\documentclass[12pt,showpacs,preprintnumbers,superscriptaddress,amsmath,amssymb,nofootinbib]{revtex4}
\usepackage{graphicx}
\usepackage{dcolumn}
\usepackage{bm}
\usepackage{amssymb}
\usepackage{amsmath}
\usepackage{epsfig}
\usepackage{color}
\usepackage{hhline}

\def\be{\begin{equation}}
\def\ee{\end{equation}}
\newcommand{\bea}{\begin{eqnarray}}
\newcommand{\eea}{\end{eqnarray}}
\newcommand{\nn}{\nonumber}

\numberwithin{equation}{section}

\begin{document}

\title{Radiative dark matter and neutrino masses\\
from an alternative $U(1)_{B-L}$ gauge symmetry}
 {\begin{flushright}{ APCTP Pre2021 - 003}\end{flushright}}
\author{Hiroshi Okada}
\email{hiroshi.okada@apctp.org}
\affiliation{Asia Pacific Center for Theoretical Physics (APCTP) - Headquarters San 31, Hyoja-dong,
Nam-gu, Pohang 790-784, Korea}
\affiliation{Department of Physics, Pohang University of Science and Technology, Pohang 37673, Republic of Korea}

\author{Yuta Orikasa}
\email{Yuta.Orikasa@utef.cvut.cz}
\affiliation{Institute of Experimental and Applied Physics, Czech Technical University in Prague, 110 00 Prague 1, Czech Republic}

\author{Yutaro Shoji}
\email{yutaro.shoji@mail.huji.ac.il}
\affiliation{Racah Institute of Physics, Hebrew University of Jerusalem, Jerusalem 91904, Israel}

\date{\today}

\begin{abstract}
We propose a model where the masses of the active neutrinos and a dark matter candidate are generated radiatively through the $U(1)_{B-L}$ gauge symmetry breaking. It is realized by a non-universal $U(1)_{B-L}$ charge assignment on the right handed neutrinos and one of them becomes DM. The dark matter mass becomes generally small compared with the typical mass of the Weak Interacting Massive Particles and we have milder constraints on the dark matter.  We consider the case where the dark matter is produced through the freeze-in mechanism and show that the observed dark matter relic density can be realized consistently with the current experimental constraints on the neutrino masses and the lepton flavor structure.

\end{abstract}
\maketitle
\newpage

\section{Introduction}
The nature of neutrinos and dark matter is one of the most attracting puzzles in particle physics and cosmology.
While we have hints of the scale of the neutrino masses, the dark matter (DM) mass is yet an arbitrary parameter ranging from $10^{-24}~{\rm eV}$ to  $10^{19}~{\rm eV}$. Huge effort has been made by experimentalists  especially on the weakly interacting massive particles (WIMP), {\it e.g.}, direct searches such as XENON1T~\cite{Aprile:2018dbl} and LUX~\cite{Akerib:2016vxi}, indirect searches such as Fermi-LAT~\cite{FermiLAT:2011ab}, AMS-02~\cite{Aguilar:2013qda}, and CALET~\cite{Adriani:2017efm}, and collider searches at LHC~\cite{Khachatryan:2014rra}.
Even though WIMPs have not been excluded, it would be better to consider lighter DM scenarios where the mass and the charges of DM are realized naturally. Some of the famous light DM scenarios are the axion DM~\cite{Preskill:1982cy} and hot/warm DM~\cite{Dodelson:1993je}. In this paper, we consider the possibility that the dark matter is one of the right handed neutrinos and obtain a small mass only radiatively in association with the radiative generation of the active neutrino masses.
Since the active neutrino masses and the DM mass are both small but non-zero, it would be natural that they have the same origin. In addition, the DM is automatically neutral under the SM gauge symmetry because it is one of the right handed neutrinos.

We consider the radiative mass generation mechanism \cite{Ma:2006km} for both of DM and the neutrinos as in \cite{Kanemura:2011vm,Baek:2013fsa, Kajiyama:2013rla}, but with a different assignment of $U(1)_{B-L}$ charges on the right handed neutrinos.
From the anomaly free condition, the three right-handed neutrinos can have non-universal charges of $(-4,-4,5)$, whose applications are found in refs.~\cite{Montero:2007cd, Sanchez-Vega:2015qva, Ma:2015mjd, Nomura:2017kih, Nomura:2017jxb, Nomura:2017vzp, Geng:2017foe, Singirala:2017cch, Okada:2018tgy, Das:2019fee, Mahapatra:2020dgk, Asai:2020xnz}. Since it is different from the ordinary universal $U(1)_{B-L}$ charge assignment, we call it as the alternative $U(1)_{B-L}$ symmetry.
Thanks to this unique charge assignment, only two right-handed neutrinos with charge $-4$ can contribute to the masses of neutrinos, while the remaining one with charge $5$ can become a DM candidate. Interestingly, one of the neutrino masses is vanishing in a minimal setup.
To generate the masses of the active neutrinos and DM radiatively, we introduce several bosons that have different charges of the $U(1)_{B-L}$ symmetry.
We also check that our model is consistent with the lepton flavor violations (LFVs), and the current DM relic abundance.

This paper is organized as follows.
In Sec.~II, we review our model and formulate the neutrino masses, LFVs, DM mass, and DM relic abundance.
Then, we show two benchmark points with DM mass of 1 MeV and 100 MeV.
In Sec.~III, we conclude.

\section{The Model}

 \begin{widetext}
\begin{center} 
\begin{table}[t]
\begin{tabular}{|c||c|c||c|c|c|}\hline\hline  
Fermions& ~$L_{L_a}$~ & ~$e_{R_a}$~ & ~$S$~ & ~$X_R$~ & ~$N_{R_i}$~
\\\hline 
 $SU(2)_L$ & $\bm{2}$ & $\bm{1}$  & $\bm{1}$  & $\bm{1}$ & $\bm{1}$   \\\hline 
$U(1)_Y$ & $-\frac12$  & $-1$  & $0$  & $0$ & $0$     \\\hline
 $U(1)_{B-L}$ & $-1$  & $-1$ & $8$   & $5$   & $-4$    \\\hline
\end{tabular}
\caption{Field contents of fermions and their charge assignment under $SU(2)_L\times U(1)_Y\times U(1)_{B-L}$, where the lower indices $a(=1-3)$ and $i(=1, 2)$ are the number of flavors.}
\label{tab:1}
\end{table}
\end{center}
\end{widetext}

\begin{table}[t]
\centering {\fontsize{10}{12}
\begin{tabular}{|c||c|c|c|c|c|c|c|}\hline\hline
  Bosons  &~$H$ ~&~ $\eta$ ~&~ $\chi$ ~&~ $\chi'$ ~&~ $\varphi$ ~&~ $\varphi'$~&~ $\varphi''$~ \\\hline
$SU(2)_L$ & $\bm{2}$ & $\bm{2}$  & $\bm{1}$ & $\bm{1}$ & $\bm{1}$ & $\bm{1}$ & $\bm{1}$  \\\hline 
$U(1)_Y$ & $\frac12$ & $\frac12$  & $0$ & $0$ & $0$& $0$ & $0$   \\\hline
 $U(1)_{B-L}$ & $0$ & $-3$ & $3$ & $13$  & $2$ & $8$ & $6$ \\\hline
\end{tabular}%
} 
\caption{Field contents of bosons and their charge assignment under $SU(2)_L\times U(1)_Y\times U(1)_{B-L}$. }
\label{tab:2}
\end{table}

In this section, we  review our scenario.
In the fermionic sector, we introduce one Dirac fermion $S$ with $8$ $B-L$ charge, one Majorana fermion $X_R$ with $5$ $B-L$ charge, which is assumed to be a DM candidate,
and two Majorana fermions $N_{R}$ with $-4$ $B-L$ charges, which are the source for the generation of the active neutrino masses and their oscillations.
The fermion field contents and their assignments are summarized in table~\ref{tab:1}.
In the bosonic sector, we introduce an inert isospin doublet boson $\eta$ with $-3$ $B-L$ charge, two inert isospin singlet boson $\chi(\chi')$ with $3(13)$ $B-L$ charges, and $\varphi(\varphi', \varphi'')$  with $2(8, 6)$ $B-L$ charges having nonzero vacuum expectation values (VEVs), denoted by  $\langle\varphi(\varphi', \varphi'')\rangle\equiv v_{\varphi(\varphi', \varphi'')}/\sqrt2$. Here $H$ is supposed to be the SM-like Higgs.
 The boson field contents and their assignments are summarized in table~\ref{tab:2}.

The relevant  Lagrangian for Yukawa sector and nontrivial scalar potential under these assignments
are given by
\begin{align}
-\mathcal{L}_{Y}
&=
y_{\ell_a} \bar L_{La} H e_{Ra} + y_{\eta_{ai}} \bar L_{L_a} \tilde\eta   N_{R_i}
+y_{N_i} \bar N^c_{R_i} N_{R_i}  \varphi' \nn\\
&+ y_{\chi'} \bar X^c_{R} S_R \chi'^* +  y_\chi \bar X_{R} S_L \chi^* + M_S \bar S_L S_R
+\rm{h.c.} \label{Lag:Yukawa}\\ 
\mathcal{V}
&\supset
 \left(\mu_0 H^\dag \eta\chi + \mu_1 \varphi \varphi'^* \varphi'' + \mu_2 \varphi'' \chi^{* 2} +{\rm h.c.} \right)  \label{eq:HP} \\
& + \left(\lambda_0 \chi \chi'^* \varphi \varphi' + \lambda_1 \chi^2 \varphi \varphi'^* 
 + \lambda_2 H^\dag \eta \chi^* \varphi'' +\lambda_3 \varphi^3\varphi''^* +\lambda_4 \chi^* \varphi'^* \varphi'^2 + {\rm h.c.}\right)
 + \lambda_5 |H^\dag \eta|,\nn 
\end{align}
where each of the index $a(=1-3)$ and $i(=1,2)$ represents the number of families,  and the first term of $\mathcal{L}_{Y}$ generates the masses
of the SM charged-lepton fermions. Here we assume all the parameters above are positive real for simplicity.\\

{\it Higgs sector}:
Here we formulate the Higgs sector.
First of all, we decompose the fields as follows:
 \begin{align}
 &H =\left[
\begin{array}{c}
w^+\\
\frac{v_H+h+iz}{\sqrt2}
\end{array}\right],\
\eta =\left[
\begin{array}{c}
\eta^+\\
\frac{\eta_R+i\eta_I}{\sqrt2}
\end{array}\right],
\varphi^{(', '')}\equiv \frac{v_{\varphi^{(', '')}}+\rho^{(', '')}+iz_{\varphi^{(', '')}} }{\sqrt2},\
\chi^{(')}\equiv \frac{\chi^{(')}_R+i\chi^{(')}_I }{\sqrt2},
\label{component}
 \end{align}
 where $z$ and $w^+$ are eaten by the SM vector boson $Z$ and $W^+$, respectively. We have one massless state after diagonalizing the mass matrix of $(z_\varphi,z_{\varphi'},z_{\varphi''})$
and it is eaten by the $B-L$ neutral vector boson $Z'$.
Then, each of the mass matrices are denoted as follows: $M_R^2$ for CP-even mass matrix in basis of $[h,\rho,\rho',\rho'']^T$,
  $M_I^2$ for CP-odd mass matrix in basis of $[z_\varphi,z_{\varphi'} ,z_{\varphi''}]^T$, $M'^2_{R}$ for inert CP-even mass matrix in basis of $[\eta_R,\chi_R,\chi_R']^T$,
  $M'^2_{I}$ for inert CP-odd mass matrix  in basis of $[\eta_I,\chi_I,\chi_I']^T$.
They are diagonalized as $O_A M^{2}_A O_A^T$ and  $V_A M'^{2}_A V_A^T$, where $A\equiv R,I$.
We denote the mass eigenstates as $[h, \rho, \rho', \rho'']^T\equiv O_R^T [h_1, h_2, h_3, h_4]^T$, $[z_\varphi, z_{\varphi'}, z_{\varphi''}]^T\equiv O_I^T [a_1, a_2, a_3]^T$,
$[\eta_R,\chi_R,\chi_R']^T\equiv V_R^T [H_1, H_2,H_3]^T$, $[\eta_I,\chi_I,\chi_I']^T\equiv V_I^T [A_1, A_2,A_3]^T$, where
$h_{\rm SM}\equiv h_1$ is the SM-like Higgs boson, and $a_1$ is eaten by the $Z'$ boson. 
In the following, we take into account the constraints on the oblique parameters and simply impose $m_{\eta^\pm}\in m_{\eta_I}\pm 120$ GeV (from $\Delta T$) and 500 GeV$\lesssim m_{\eta_R}\approx m_{\eta_I}$  (from $\Delta S$)~\cite{Barbieri:2006dq}.  
\\

{\it Active neutrinos}:
Here we formulate the active neutrino sector.
We start with the Lagrangian written in terms of mass eigenstate:
\begin{align}
-{\cal L}_\nu&=\frac{y_{\eta_{ai}}(V^T_R)_{1\alpha} }{\sqrt2} \bar \nu_{L_a} N_{R_i} H_\alpha
-i \frac{y_{\eta_{ai}}(V^T_I)_{1\alpha} }{\sqrt2} \bar \nu_{L_a} N_{R_i} A_\alpha,
\end{align}
where $\alpha (=1-3)$ should be summed up.
Then the  active neutrino mass matrix, $m_\nu$,
is given at the one-loop level via three inert bosons, and its formula is given by
\begin{align}
(m_{\nu})_{ab}
 &= -\frac{1}{2 (4\pi)^2}\sum_{\alpha =1}^3
  \sum_{i=1}^2 y_{\eta_{ai}} M_i y^T_{\eta_{ib}}
  \left( (V_R )_{1\alpha}^2F_I[H_\alpha, i]
  - (V_I )_{1\alpha}^2F_I[A_\alpha, i]
  \right),
\\
 F_I[a, i]
 &= \frac{m_a^2}{M_i^2-m_a^2} \ln\left( \frac{m_a^2}{M_i} \right),
\end{align}
where $M_i\equiv v_{\varphi'}y_{N_i}/\sqrt2$. Here $m_\nu$ is diagonalized by the neutrino mixing matrix, $V_{\rm MNS}$, as 
$m_\nu  = V_{\rm MNS}^\dag D_\nu V_{\rm MNS}^*~(D_\nu\equiv V_{\rm MNS} m_\nu V^T_{\rm MNS})$ with $D_\nu\equiv 
(m_{\nu_1},m_{\nu_2},m_{\nu_3})$ with $m_{\nu_1}(m_{\nu_3})=0$ for normal(inverted) ordering.
Then we can parameterize the Yukawa coupling in terms of an arbitrary complex $3\times2$ rotation matrix
with $O^TO=1_{2\times2}$, as 
follows
\begin{align}
& y_\eta = V_{\rm MNS}^\dag \sqrt{D_\nu} O R^{-1/2},
\\
&R = -\frac{1}{2 (4\pi)^2}
  \sum_{\alpha=1}^3  M_i
  \left( (V_R )_{1\alpha}^2F_I[H_\alpha, i]
  - (V_I )_{1\alpha}^2F_I[A_\alpha, i]
  \right),
\label{diagR}
\end{align}
where $m_\nu\equiv y_\eta R y^T_\eta$. 
And $O$ for the normal hierarchy (NH) and the inverted hierarchy (IH) are given by
\begin{align}
\left[\begin{array}{cc}
0 & 0\\
\cos \theta & -\sin \theta \\
\pm \sin \theta & \pm\cos \theta \\
\end{array}\right], \quad 
\left[\begin{array}{cc}
\cos \theta & -\sin \theta \\
\pm \sin \theta & \pm\cos \theta \\
0 & 0\\
\end{array}\right],  
\label{eq:omix}
\end{align}  
respectively. Notice that $\theta$ can be complex. 
We assume the perturbative bound; $y_\eta\lesssim \sqrt{4\pi}$.
This parameterization allows us to use the neutrino oscillation data as input parameters in our numerical analysis. We use the values in NuFIT 5.0~\cite{Esteban:2020cvm}.

{\it Lepton flavor violations (LFVs)}:
LFV processes $\ell_a \to \ell_b \gamma$ arise at the one-loop level from the same Yukawa couplings used for the generation of the neutrino masses, and its formula is given by~\cite{Lindner:2016bgg, Baek:2016kud}
\begin{align}
BR(\ell_a\to \ell_b \gamma)&\approx\frac{\pi^3\alpha_{em}C_{\alpha\beta}}{3(4\pi)^4G_F^2}
\left|\sum_{i=1}^2 (y_\eta)_{ai} (y_\eta^\dag)_{ib} F_{lfv}(i,\eta^\pm)\right|^2,\\
F_{lfv}(a,b)&\equiv\frac{2 m_a^6+3m_a^4m_b^2-6m_a^2m_b^4+m_b^6+12m_a^4m_b^2\ln\left[\frac{m_b}{m_a}\right]}{(m_a^2-m_b^2)^4},
\end{align}
where $\alpha_{em}\approx1/137$ is the fine-structure constant, $G_F\approx1.17\times10^{-5}$ GeV$^{-2}$ is the Fermi constant,
and $C_{21}\approx1$, $C_{31}\approx 0.1784$, $C_{32}\approx0.1736$. 
The experimental upper bounds are found in~\cite{TheMEG:2016wtm, Adam:2013mnn}: 
\begin{equation}
{\rm BR}(\mu\to e \gamma)\lesssim 4.2\times 10^{-13},\ 
{\rm BR}(\tau\to e \gamma)\lesssim 3.3\times 10^{-8},\ 
{\rm BR}(\tau\to \mu \gamma)\lesssim 4.4\times 10^{-8},
 \end{equation}
where we define $\ell_1\equiv e$,  $\ell_2\equiv \mu$, and  $\ell_3\equiv \tau$.

We estimate the size of LFVs in our model. 
We assume the scalar masses are the same order ($\sim m_S$) and the fermion masses are same order ($\sim M_f$).
In the case of $r \equiv \frac{m_s}{M_f} \geq 1$, the Yukawa couplings and the branching ratios are written as 
\begin{eqnarray}
y_\eta &\sim& {\sqrt \frac{D_\nu}{M_f F_I[r, 1]}}, 
\\
BR(\ell_a\to \ell_b \gamma)&\sim& \frac{\pi^3\alpha_{em}C_{\alpha\beta}}{3(4\pi)^4G_F^2}
\left|  \frac{D_\nu}{M_f^3 F_I[r, 1]}  F_{lfv}(1, r)\right|^2
\nonumber\\
&\sim& 10^{-36} \left( \frac{1 \ {\rm TeV}}{M_f} \right)^6 \left(  \frac{ F_{lfv}(1, r)}{F_I[r, 1]}\right)^2, 
\end{eqnarray} 
where 
$\left| F_{lfv}(1, r)/F_I[r, 1] \right| \leq \frac12$ for $r \geq 1$. 
In the case of $r<1$, we use the following expression: 
\begin{eqnarray}
y_\eta &\sim& {\sqrt \frac{r D_\nu}{ m_s F_I[1, 1/r]}}, 
\\
BR(\ell_a\to \ell_b \gamma)&\sim& \frac{\pi^3\alpha_{em}C_{\alpha\beta}}{3(4\pi)^4G_F^2}
\left|  \frac{r D_\nu}{m_s^3 F_I[1, 1/r]}  F_{lfv}(1/r, 1)\right|^2
\nonumber\\
&\sim& 10^{-36} \left( \frac{1 \  {\rm TeV}}{m_s} \right)^6 \left(  \frac{r F_{lfv}(1/r, 1)}{F_I[1, 1/r]}\right)^2, 
\end{eqnarray}
where $\left| r F_{lfv}(1/r, 1)/F_I[1, 1/r] \right| < \frac12$ for $r<1$.
In both cases, the branching ratios are much smaller than the experimental bounds. 
We can obtain ${\cal O}(1)$ Yukawa couplings using the complex phases in the complex orthogonal matrix and 
a cancellation in Eq.(\ref{diagR}).
However, since these situations need a fine tuning or a hierarchical structure in the scalar mass matrices, we do not consider the situation in this paper.

{\it The warm dark matter candidate}:
Here, we derive the DM mass $M_X$ at one-loop level.
We first write the relevant Lagrangian in terms of mass eigenstate as follows:
\begin{align}
-{\cal L}_X&=
\frac{y_{\chi'} (V^T_R)_{3\alpha} }{\sqrt2} \bar S_R X^C_R H_\alpha
+
\frac{y_{\chi} (V^T_R)_{2\alpha} }{\sqrt2} \bar X_R S_L H_\alpha 
-i\frac{y_{\chi'} (V^T_I)_{3\alpha'} }{\sqrt2} \bar S_R X^C_R A_{\alpha'}
-
i\frac{y_{\chi} (V^T_I)_{2\alpha'} }{\sqrt2} \bar X_R S_L A_{\alpha'} ,
\end{align}
where $\alpha,\alpha'(=1-3)$ should be summed up.
Then the  DM mass is given by
\begin{align}
M_X
&=
-\frac{ y_\chi  M_S y_{\chi'}}{(4\pi)^2}
 \sum_{\alpha=1}^3 
 \left[
  (V^T_R)_{3\alpha} (V^T_R)_{2\alpha} F_I[S, H_\alpha] 
  - (V^T_I)_{3\alpha} (V^T_I)_{2\alpha} F_I[S, A_\alpha] 
   \right] .
\end{align}
\\

{\it Relic density}:
We consider the case where the DM is $X$ and is generated via the freeze-in mechanism. In order to simplify our discussion, we consider a reheating temperature that is much lower than the $W/Z$ masses and the scalar masses but much higher than the $Z'$ mass so that the main production processes are through the $Z'$ boson. Notice that the $Z'$ mass becomes small due to the small gauge coupling required to realize the correct DM abundance. We assume that the initial abundance of $Z'$ and that of $X$ are initially zero and that $Z'$ and $X$ are in kinetic equilibrium with the SM particles. We calculate their current abundance by solving the Boltzmann equations given below.

Since we consider a rather light DM, which can also be produced after the QCD phase transition, we use different Boltzmann equations above and below the transition temperature, $T_{\rm QCD}$.
For $T>T_{\rm QCD}$, they are given by~\footnote{Similar analysis has been done by, e.g., ref.~\cite{Kaneta:2016vkq}.}
\begin{align}
\frac{dY_{Z'}}{dx}&=c(x)\left[\sum_f\mathcal C(f\bar f\to Z')+\mathcal C(XX\to Z')\right],\label{boltz_1}\\
\frac{dY_{X}}{dx}&=2c(x)\left[\mathcal C(Z'\to XX)+\sum_f\mathcal C(f\bar f\to Z^*\to XX)\right],\label{boltz_2}
\end{align}
where $Y_{Z'}$ and $Y_X$ are the yields of $Z'$ and $X$, respectively, and the sum is taken over all the SM leptons and quarks. The collision terms, $\mathcal C$'s, are given in Appendix \ref{apx:collision}.  We have defined
\begin{align}
    x&=\frac{m_X}{T},\\
    c(x)&=x^4\frac{135\sqrt{10}}{2\pi^3}\frac{m_{\rm Pl}}{m_X^5}\frac{1}{g_*^{1/2}g_{*s}}\left(1+\frac{1}{3}\frac{d\ln g_{*s}}{d\ln T}\right).
\end{align}
Here, $g_*$ is the temperature dependence of the effective degrees of freedom for the energy density and  $g_{*s}$ is those for the entropy. Their evolution is taken from \cite{Drees:2015exa}.

For $T<T_{\rm QCD}$, Eq.~\eqref{boltz_2} is modified as
\begin{align}
\frac{dY_{X}}{dx}&=2c(x)\left[\mathcal C(Z'\to XX)+\sum_\ell\mathcal C(\ell\bar \ell\to Z^*\to XX)\right].
\end{align}
Here, the sum of $\ell$ is taken over the SM leptons.
For Eq.~\eqref{boltz_1}, we modify if $m_{Z'}\lesssim1~{\rm GeV}$ is satisfied since otherwise we cannot rely on the effective theory of hadrons. We use
\begin{align}
\frac{dY_{Z'}}{dx}&=c(x)\left[\sum_\ell\mathcal C(\ell\bar \ell\to Z')+\mathcal C(XX\to Z')+\mathcal C(\pi^0\gamma\to Z')+\mathcal C(\pi^0\pi^+\pi^-\to Z')\right].
\end{align}
We ignore $\pi^0\gamma\to XX$ and $\pi^0\pi^+\pi^-\to XX$ processes via the off-shell $Z'$ since the pions disappear soon after the QCD phase transition and will not affect the abundance significantly. The other hadronic decay channels are known to be smaller than the above two processes \cite{Tulin:2014tya}.

We solve the Boltzmann equation from $x=M_X/T_R$ with $T_R$ being the reheating temperature, to a sufficiently large $x=x_{\infty}$. The dark matter relic density should satisfy \cite{Aghanim:2018eyx}
\begin{equation}
    \Omega_Xh^2=\frac{Y_X(x_{\infty})s_0M_X}{3M_{\rm Pl}^2H_{100}^2}=0.1193 \pm 0.0018,
\end{equation}
where $s_0$ is the current entropy density, $H_{100}=100~{\rm km/s/Mpc}$.

Since the gauge coupling and the $Z'$ mass are independent of the constraints discussed in the previous sections, we can always tune them to obtain the correct relic abundance. Thus, we only show two distinct parameter sets that give $\Omega h^2\simeq0.12$.

\begin{figure}[t]
    \begin{minipage}{0.48\linewidth}
        \centering
        \includegraphics[width=\linewidth]{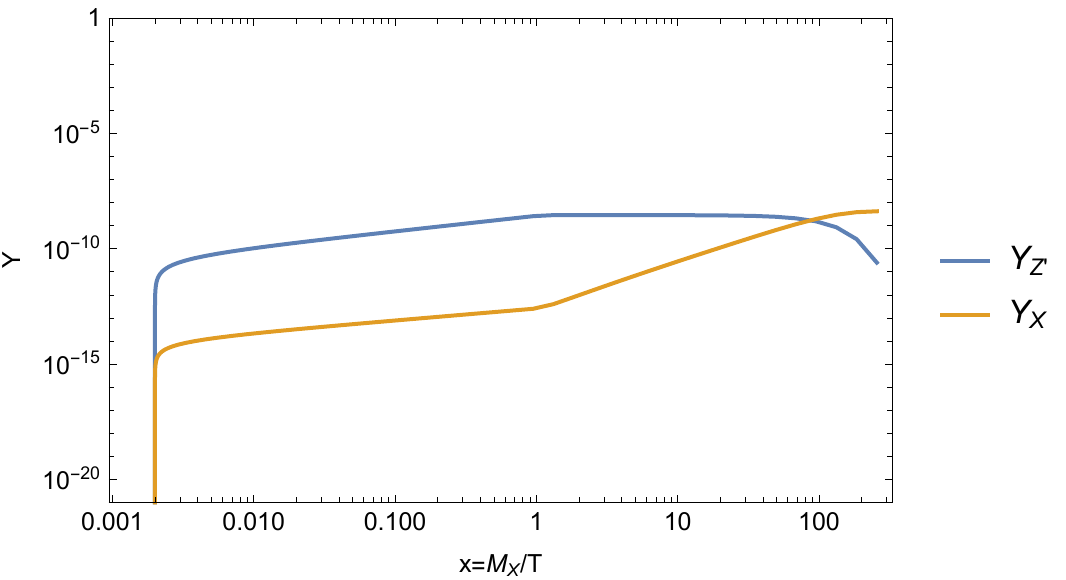}
    \end{minipage}
    \begin{minipage}{0.48\linewidth}
        \centering
        \includegraphics[width=\linewidth]{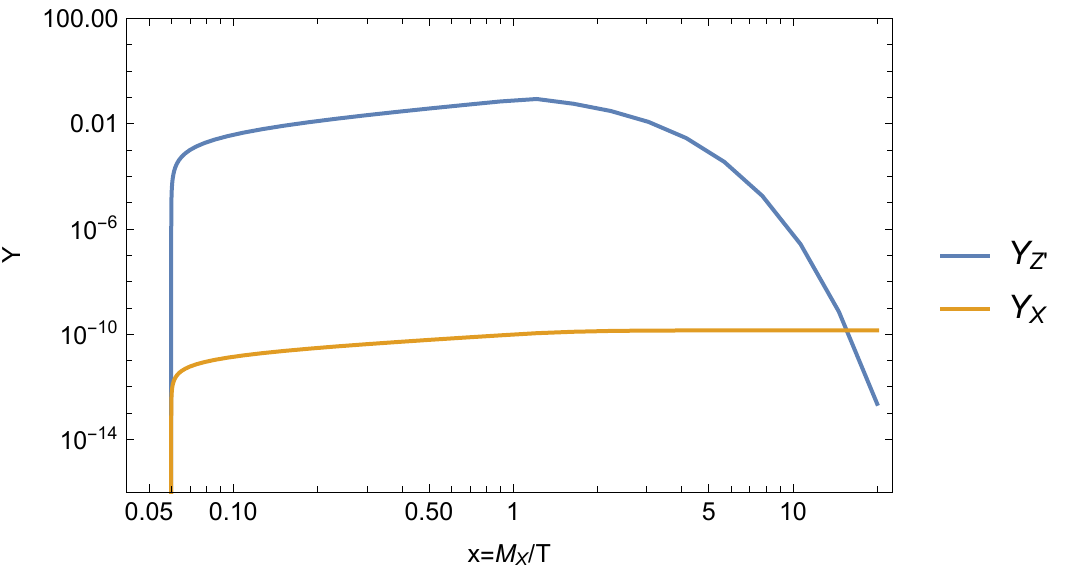}
    \end{minipage}
    \caption{The evolution of the yields. The left panel is for Eq.~\eqref{eq:param1} and the right panel is for Eq.~\eqref{eq:param2}.}
    \label{fig:yields}
\end{figure}

In Fig.~\ref{fig:yields}, we plot the evolution of the yields of the DM and the mediator.
In the left panel, we take
\begin{align}
    M_X=100~{\rm MeV},~m_{Z'}=5~{\rm GeV},~g'=2.3\times10^{-12},~T_R=50~{\rm GeV}.\label{eq:param1}
\end{align}
With this parameter set, the mediator can decay into DM and the gauge coupling is required to be very small. The large part of the DM relic density is coming from the decay of the mediator at $1\lesssim x\lesssim100$.

In the right panel, we take
\begin{align}
    M_X=3~{\rm GeV},~m_{Z'}=5~{\rm GeV},~g'=1\times10^{-6},~T_R=50~{\rm GeV}.\label{eq:param2}
\end{align}
With this parameter set, DM is generated only through off-shell mediator and thus we need a rather large gauge coupling. Since the mediator cannot decay into a DM pair, the large yield of $Z'$ does not affect the DM relic density.

For these parameter sets, the constraints from the beam dump experiments \cite{Bross:1989mp, Riordan:1987aw, Davier:1989wz, Blumlein:2013cua, Bjorken:1988as, Seto:2020udg} and SN1987A \cite{Dent:2012mx, Kazanas:2014mca} can be evaded.

\section{Conclusions}
We have proposed a model where the active neutrino masses are generated radiatively and one of the right handed neutrinos becomes DM. We have naturally realized the tiny mass of DM as well as the neutrino mass matrix in the successful framework of the alternative gauged $U(1)_{B-L}$ symmetry. 
We have shown that the LFV constraints are typically weak for this model.
We have also investigated the relic abundance of DM through the freeze-in mechanism via the $Z'$ gauge boson mediation. 
We have shown two successful benchmark points that realize the correct relic density.

\section*{Acknowledgments}
\vspace{0.5cm}
The work was supported by the Junior Research Group (JRG) Program at the Asia-Pacific Center for Theoretical
Physics (APCTP) through the Science and Technology Promotion Fund and Lottery Fund of the Korean Government, 
the Korean Local Governments-Gyeongsangbuk-do Province and Pohang City (H.O.)
and European Regional Development Fund-Project Engineering Applications of Microworld Physics 
(No. CZ.02.1.01/0.0/0.0/16\_019/0000766)(Y.O.).
H.O.~is sincerely grateful for all the KIAS members. Y.S.~is supported by  I-CORE Program of the
Israel Planning Budgeting Committee (grant No. 1937/12). The authors would like to thank Dr.~Osamu Seto for his useful comments.

\appendix
\section{Collision terms}
\label{apx:collision}
The collision terms for $1\leftrightarrow2$ processes are given by
\begin{align}
\mathcal C(f\bar f\to Z')=\frac{N_fQ_f^2g'^2m_{Z'}^3T}{8\pi^3}\left(1-\frac{Y_{Z'}}{Y_{Z'}^{{\rm eq}}}\right)\left(1+\frac{2m_f^2}{m_{Z'}^2}\right)\beta_f(m_{Z'}^2)K_1\left(\frac{m_{Z'}}{T}\right),
\end{align}
and
\begin{align}
\mathcal C(Z'\to XX)=-\mathcal C(XX\to Z')=\frac{Q_X^2g'^2m_{Z'}^3T}{16\pi^3}\left(\frac{Y_{Z'}}{Y_{Z'}^{{\rm eq}}}-\frac{Y_{X}^2}{(Y_{X}^{{\rm eq}})^2}\right)\beta_X^3(m_{Z'}^2)K_1\left(\frac{m_{Z'}}{T}\right),
\end{align}
where $N_f$ is $3$ for the quarks, $1$ for the charged leptons, and $1/2$ for the neutrinos. Here,
\begin{equation}
    \beta_f(s)=\Re\sqrt{1-\frac{4m_f^2}{s}},~\beta_X(s)=\Re\sqrt{1-\frac{4m_X^2}{s}}.
\end{equation}

For $f\bar f\to XX$ process, we need to take care of the double counting. The sum of the on-shell and the off-shell $Z'$ contributions is calculated as
\begin{align}
\mathcal C(f\bar f\to XX)&=\frac{N_fQ_X^2Q_f^2g'^4T}{192\pi^5}\left(1-\frac{Y_{X}^2}{(Y_{X}^{{\rm eq}})^2}\right)\int ds\frac{s^{5/2}\beta_X^3(s)\beta_f(s)}{|s-m_{Z'}^2+im_{Z'}\Gamma_{Z'}|^2}\left(1+\frac{2m_f^2}{s}\right)K_1\left(\frac{\sqrt{s}}{T}\right).
\end{align}
Since the on-shell part is already taken into account by the Boltzmann equation for $Z'$, we eliminate it as
\begin{align}
    \mathcal C(f\bar f\to Z^*\to XX)&=\mathcal C(f\bar f\to XX)\nonumber\\
    &\hspace{3ex}-\mathcal C(f\bar f\to Z'){\rm Br}(Z'\to XX)\left(1-\frac{Y_{X}^2}{(Y_{X}^{{\rm eq}})^2}\right)\left(1-\frac{Y_{Z'}}{Y_{Z'}^{{\rm eq}}}\right)^{-1}.
\end{align}
Here, the decay width of the $Z'$ boson is given by
\begin{equation}
    \Gamma_Z=\sum_f\Gamma(Z'\to f\bar f)+\Gamma(Z'\to XX),
\end{equation}
with
\begin{align}
    \Gamma(Z'\to f\bar f)&=\frac{N_fQ_f^2g'^2m_{Z'}}{12\pi}\left(1+\frac{2m_f^2}{m_{Z'}^2}\right)\beta_f(m_{Z'}^2),\\
    \Gamma(Z'\to XX)&=\frac{Q_X^2g'^2m_{Z'}}{24\pi}\beta_X^3(m_{Z'}^2).
\end{align}

The collision terms involving pions are \cite{Tulin:2014tya}
\begin{align}
    \mathcal C(\pi^0\gamma\to Z')&=\frac{\alpha_{\rm EM}g'^2m_{Z'}^5T}{256\pi^6f_\pi^2}\left(1-\frac{Y_{Z'}}{Y_{Z'}^{\rm eq}}\right)\left(1-\frac{m_\pi^2}{m_{Z'}^2}\right)^3K_1\left(\frac{\sqrt{s}}{T}\right),\\
    \mathcal C(\pi^0\pi^+\pi^-\to Z')&=\frac{g_{\rho\pi\pi}^4g'^2m_{Z'}^3T}{512\pi^9f_\pi^2}\mathcal I(m_{Z'})\left(1-\frac{Y_{Z'}}{Y_{Z'}^{\rm eq}}\right)K_1\left(\frac{\sqrt{s}}{T}\right).
\end{align}
Here, $g_{\rho\pi\pi}^2/(4\pi)\simeq3$ and $\mathcal I(m_{Z'})$ is the phase space integral given in \cite{Tulin:2014tya}. We consider them only when they are kinematically allowed.
Similarly, for the decay width of $Z'$, we have \cite{Tulin:2014tya}
\begin{align}
    \Gamma_Z&=\sum_\ell\Gamma(Z'\to \ell\bar \ell)+\Gamma(Z'\to XX)+\Gamma(Z'\to \pi^0\gamma)+\Gamma(Z'\to \pi^+\pi^-\pi^0),
\end{align}
with
\begin{align}
    \Gamma(Z'\to \pi^0\gamma)&=\frac{g'^2\alpha_{\rm EM}m_{Z'}^3}{384\pi^4f_\pi^2}\left(1-\frac{m_\pi^2}{m_{Z'}^2}\right)^3,\\
    \Gamma(Z'\to \pi^+\pi^-\pi^0)&=\frac{g_{\rho\pi\pi}^4g'^2m_{Z'}}{768\pi^7f_\pi^2}\mathcal I(m_{Z'}).
\end{align}

\end{document}